\documentclass[sn-mathphys-num]{sn-jnl}% Math and Physical Sciences Numbered Reference Style 
\usepackage{graphicx}%
\usepackage{multirow}%
\usepackage{amsmath,amssymb,amsfonts}%
\usepackage{amsthm}%
\usepackage{mathrsfs}%
\usepackage[title]{appendix}%
\usepackage{xcolor}%
\usepackage{textcomp}%
\usepackage{manyfoot}%
\usepackage{booktabs}%
\usepackage{algorithm}%
\usepackage{algorithmicx}%
\usepackage{algpseudocode}%
\usepackage{listings}%
\theoremstyle{thmstyleone}%
\theoremstyle{thmstyletwo}%

\theoremstyle{thmstylethree}%

\begin{document}

\title[Article Title]{Organic compounds in metallic hydrogen.}


\author[1,2]{\fnm{Jakkapat} \sur{Seeyangnok}}\email{jakkapatjtp@gmail.com}

\author[2]{\fnm{Udomsilp} \sur{Pinsook}}\email{udomsilp.p@chula.ac.th}

\author[1]{\fnm{Graeme John} \sur{Ackland}}\email{gjackland@ed.ac.uk}

\affil[1]{\orgdiv{Centre for Science at Extreme Conditions, School of Physics and Astronomy}, \orgname{University of Edinburgh},  \city{Edinburgh}, \postcode{EH9 3FD}, \state{Scotland}, \country{United Kingdom}}

\affil[2]{\orgdiv{Department of Physics, Faculty of Science}, \orgname{Chulalongkorn University}, \orgaddress{\street{254 Phaya Thai Rd}, \city{Bangkok}, \postcode{10330}, \country{Thailand}}}


\abstract{Metallic hydrogen\cite{wigner1935possibility} is the most common condensed material in the universe, comprising the centre of gas giant planets\cite{militzer2016understanding,nettelmann2012jupiter,wahl2017comparing}  However, experimental studies are extremely challenging\cite{loubeyre2020synchrotron,dias2017observation,eremets2011conductive,knudson2015direct}, and most of our understanding of this material has been led by theory.  Chemistry in this environment has not been probed experimentally, so here we examine hydrocarbon chemistry in metallic hydrogen using density functional theory calculations\cite{pickard2007structure,geng2016predicted}.   We find that carbon and oxygen react with metallic hydrogen to produce an entirely new type of hydrocarbon chemistry based on sixfold coordinated carbon with organic-style molecules CH$_6$, C$_2$H$_8$,  C$_3$H$_{10}$ OH$_3$ NH$_4$ and CH$_4$OH.  These are charged molecules stabilised  by the metallic environment. Their associated electric fields are screened, giving oscillation in the surrounding electron and proton densities.  In view of the excess hydrogen we refer to them as hypermethane, hyperethane  etc. The relationship to traditional chemistry is that the metallic background acts as an electron donor and stabilizes negatively charged ions.  This enables the formation of six covalent bonds per carbon atom, or three per oxygen atom.  This demonstrates that organic chemistry may take place in very different environments from those found on earth, and may be common throughout the universe.}




\maketitle

\section{Introduction}\label{sec1}
Metallic hydrogen is believed to be the most common condensed phase of matter in the universe, comprising the cores of gas-giant planets, and giving rise to their enormous magnetic fields. 
However, it is exceptionally challenging to make metallic hydrogen on earth.  Consequently, most of our understanding of this material comes from theory. 

Modern planetary models depict layers separated by the weight of elements, e.g. gas giants feature an outer molecular hydrogen and helium envelope and a core of  metallic hydrogen depleted of helium \cite{hubbard1980structure,guillot1999comparison,helled2018interiors,wahl2017comparing} in
which is predicted to be insoluble below its own metallization conditions
\cite{militzer2013ab,lorenzen2009demixing,preising2020metallization,schottler2018ab,militzer2013equation,clay2016benchmarking,ramsey2020localization,soubiran2017properties}.  Understanding of material properties allows us to infer the composition and structure of exoplanets from their mass-radius relation\cite{swift2011mass,militzer2013ab,militzer2016understanding,movshovitz2020saturn,nettelmann2008ab,nettelmann2012jupiter,french2012ab,holst2008thermophysical}.
Chemical bonding is different at high pressure. For example, on earth, the major components of the mantle exhibit sixfold coordinated silicon\cite{tschauner2014discovery,liu1976high,yagi1978structure,ito1978synthesis,murakami2004post}, in contrast to the fourfold sp$^3$ bonding found in normal conditions.   The unconventional formation of sixfold coordinated silicon is well described by density functional theory calculations as being enabled by the electrons donated from the Mg ions, and stabilised at pressure thanks to the increased density\cite{wentzcovitch1995ab,warren1996ab,karki1997elastic,warren1998phase,wentzcovitch2006mgsio3,tsuchiya2005vibrational,umemoto2017phase}.  If chemistry can change so radically just few thousand kilometres beneath our feet, how different might it be elsewhere in the solar system?

While helium in hydrogen, and high pressure hydrogen-rich metals are very well studied, particularly with potential applications to superconductivity, less attention has been paid to the issue of solubility of heavier elements in metallic hydrogen, and the implications different chemical bonding within giant planets\cite{wilson2012rocky,soubiran2017properties,zhang2018path,roy2024mixture}.
    
Theoretical study of metallic hydrogen began in 1935, when Wigner and Huntington \cite{wigner1935possibility} used free electron theory to estimate the density of metallic hydrogen, obtaining a value remarkably close to current estimates. This implied that  hydrogen molecules would transition into atomic metallic hydrogen when subjected to sufficiently high pressure - unfortunately their estimate of 25GPa was more than an order of magnitude too low.  In 1968, Ashcroft made a remarkable prediction that metallic hydrogen would be a room temperature superconductor.  Predictions of the crystal structure of metallic hydrogen came even later\cite{pickard2007structure,mcmahon2011ground,pickard2012density,monserrat2018structure}, through ab initio random-structure exploration.  
Theoretical assessments \cite{mcmahon2011ground, pickard2007structure, johnson2000structure,clay2014benchmarking,azadi2017role,magduau2013identification,magduau2017simple} indicate that at low temperature hydrogen remains molecular to about 500GPa. 
 Surprisingly, the first atomic and metallic phase  is now believed to be a complex, open structure, rather than the dense-packed structures assumed by Wigner, Huntingdon and Ashcroft. This type of open structure is typical of the high pressure Group I electride materials, having $I4_{1}/amd$ symmetry,  isostructural with  cesium IV\cite{mcmahon2006high,falconi2006ab,tse2020chemical,miao2014high}. 500GPa is at the limit of current experiments which have seen signs of bandgap closure and reflectivity.\cite{loubeyre2020synchrotron,dias2017observation,eremets2011conductive}. This  $I4_{1}/amd$ structure persists until 2.5 TPa, beyond which more densely packed structures are favoured\cite{geng2012high,mcmahon2011ground}.  Fluid metallic hydrogen has been detected at much lower pressures, but higher temperatures, in both static and dynamic compression\cite{knudson2015direct,zaghoo2017conductivity}, and calculation\cite{nettelmann2008ab,geng2019thermodynamic,van2021isotope,lue2024re}.
    
% As pressure nears 500 GPa, hydrogen molecules disintegrate into a monoatomic body-centered tetragonal arrangement within the $I4_{1}/amd$ space group, where the ratio $c/a$ exceeds 1, in line with predictions from \cite{mcmahon2011ground, pickard2007structure}. While \cite{mcmahon2011ground} acknowledges the possibility of a scenario where $c/a<1$, this configuration rapidly loses stability under increasing pressure. Furthermore, the stability of the $I4_{1}/amd$ structure persists until 2.5 TPa, beyond which more densely packed structures are favoured.\cite{geng2012high,mcmahon2011ground}
    
In experiments\cite{gregoryanz2020everything}, synthesis of solid metallic hydrogen has been claimed at pressures exceeding 420 GPa using infrared absorption measurements \cite{loubeyre2020synchrotron}, and at an even higher pressure of 495 GPa as evidenced by reflectivity measurements \cite{dias2017observation}. Liquid metallic hydrogen has been reported at much lower pressures both in experiment \cite{nellis1998metallization,knudson2015direct,zaghoo2017conductivity} and simulation\cite{lorenzen2010first,geng2019thermodynamic,cheng2020evidence,van2021isotope}.

Carbon is particularly important, being the fourth most abundant element and the building block of organic chemistry.   The solubility of hydrocarbons in metallic hydrogen remains the preserve of theorists. Hydrocarbons have recently been studied with planetary interior conditions, i.e. high pressure and high temperature \cite{kraus2023indirect,kraus2018high,hartley2020dynamically,schuster2020measurement,huang2023formation,ranieri2022formation,conway2019high,conway2021rules,naumova2021unusual,ancilotto1997dissociation,gao2010dissociation,roy2024mixture,helled2020understanding}. These studies are useful for understanding the interior of giant planets, such as Neptune and Uranus \cite{schuster2020measurement,naumova2021unusual,ancilotto1997dissociation,gao2010dissociation} , and Jupiter and Satern \cite{helled2020understanding}. Some special compounds, such as CH$_4$(H$_2$)$_2$ \cite{ranieri2022formation,conway2019high}, C$_2$H$_6$ and C$_4$H$_{10}$ \cite{gao2010dissociation}, can be formed under these extreme conditions. In addition, there were a number of studies on the hydrocarbon \cite{roy2024mixture} and helium \cite{helled2020understanding} in the hydrogen environment.
Studies on giant planets suggest that hydrocarbons likely exist within the middle layer of their structure \cite{ancilotto1997dissociation,benedetti1999dissociation,gao2010dissociation}. Methane (CH$_{4}$) has been identified as the most abundant hydrocarbon at pressures of up to several hundred GPa and forms hydrogen-rich compounds with H$_2$ up to 160GPa\cite{hubbard1991interior,ranieri2022formation}. At higher pressures, simulations suggested that methane decomposes into hydrogen and diamond \cite{ancilotto1997dissociation,benedetti1999dissociation,roy2024mixture}. Due to its  density, the latter subsequently gravitationally sinks deeper into the planet in a phenomenon known as diamond rain.\cite{ancilotto1997dissociation,benedetti1999dissociation,ross1980repulsive,nellis1981shock,nellis2001electrical,zerr2006decomposition,kolesnikov2009methane,he2022diamond,frost2024diamond}.  This predicted demixing contrasts with the observation of high pressure reaction between diamond and hydrogen\cite{pena2021situ} - an issue which has caused significant practical challenges to synthesizing metallic hydrogen in diamond anvil cells\cite{gregoryanz2020everything}.

One challenge for theory is the richness of hydrocarbon chemistry.  The demonstration that methane is unstable to decomposition does not preclude other stable hydrocarbons.  Moreover, given the high temperatures and  excess of hydrogen over carbon in gas giant planets, even a low solubility limit could result in much of the carbon remaining in solution in the metallic hydrogen layers.

Here, we use density functional theory calculations to consider what form of carbon will exist in a metallic hydrogen environment. 
We start by examining the  free energy in the well-characterized case of solid solution carbon in crystalline I$4_1$/amd metallic hydrogen.  Then we demonstrate the equivalence of the molecular dynamics approach as a reliable estimator of thermodynamic properties, and apply molecular dynamics to investigate the planetary-relevant fluid metallic hydrogen.
     
We predict the existence of a new hydrocarbon chemistry, based around a basic sixfold coordination of carbon and threefold coordination of oxygen.  For example we observe CH$_6$,   C$_2$H$_8$,  C$_3$H$_{10}$, OH$_3$ and CH$_4$OH.

\section{Results and Discussion}
    \subsection{Solid solubility of carbon in I$4_1$/amd metallic hydrogen}

        \begin{table}[h] 
        \centering
        \begin{tabular}{ccccccc}   
       \hline
            Supercell & $H_{MD}$ & $H_{static}$ & $TS$ (MD) & $U_{ZPE}$ (MD) & $G$ (MD) \\
            \hline
            CH$_{126}$ & -10.7925$\pm$0.0006 & -10.91109 & 0.00988 & 0.32588 & -10.4764$\pm$0.0006  \\
            CH$_{125}$ & -10.8024 $\pm$0.0010 & -10.92137 & 0.00950 & 0.32125 & -10.4906$\pm$0.0010  \\
            CH$_{124}$ & -10.8139$\pm$0.0007 & -10.93038 & 0.00875 & 0.32154 & -10.5011$\pm$0.0007  \\
            CH$_{123}$ & -10.8176$\pm$0.0010 & -10.93814 & 0.01094 & 0.31637 & -10.5122$\pm$0.0010  \\
            %CH$_{121}$ & -10.83341 & -10.95343 & 0.01149 & 0.32468 & -10.52021 \\
            C (diamond) & -144.0846$\pm$0.0016 & -144.35980 & 0.00414 & 0.27343 &  -143.8153$\pm$0.0016  \\
            H (I4amd) &  -9.7593$\pm$0.0004 & -9.87520 & 0.00488 & 0.33264 & -9.4316$\pm$0.0004  \\
            \hline
            Supercell &  & $H_{static}$ & $TS$ (DFPT) & $U_{ZPE}$ (DFPT) &  \\
            \hline
            CH$_{124}$ & &-10.93038 & 0.00611 & 0.29304 & \\
            C (diamond) & & -144.3598& 0.00404 & 0.27565 &   \\
            H (I4amd) & & -9.87520 & 0.00358 & 0.29202 &  \\
            \hline
            Supercell & $\Delta H_{MD}$ & $\Delta H_{static}$  & $-T\Delta S$ (MD) & $\Delta U_{ZPE}$ (MD) & $g_{sol-MD}$ \\
            \hline
            CH$_{126}$ & 3.12$\pm$0.13 & 2.92644 & -0.63543 & -0.79825 & 1.69$\pm$0.13 \\
            CH$_{125}$ & 2.90$\pm$0.18 & 2.66684 & -0.58265 & -1.37589 & 0.94$\pm$0.18 \\
            CH$_{124}$ & 2.51$\pm$0.14 & 2.58685 & -0.48408 & -1.32760 & 0.69$\pm$0.14 \\
            CH$_{123}$ & 3.10$\pm$0.17 & 2.68081 & -0.75218 & -1.95750 & 0.39$\pm$0.17 \\
            %CH$_{121}$ & 3.28807 & 2.92644 & - 0.80661 & -0.91096 & 1.57049 \\
            \hline
        \end{tabular}
         \caption{\textbf{Thermodynamic properties for the solid solution at 300K and 500GPa} 
         As-calculated enthalpy $H$ (eV/atom), entropy $TS$ (eV/atom), zero-point energy $U_{ZPE}$ (eV/atom), configurational entropy (eV/atom), Gibbs free energy (eV/atom), and the free energy of solution (eV) for different arrangements.  Calculated energies are relative to the fully ionised atomic states, hence the high degree of precision required.  $H_{MD}$ represents the ensemble average enthalpy obtained from  NPT molecular dynamics at 500GPa, while $H_{static}$ is derived from static optimization. Entropy $TS$ is the summation of both vibrational and configurational entropy ($S_{vib}+S_{conf.}$).  $U_{ZPE}$ and $S_{vib}$ were calculated using the phonon density of states. The upper two sections are per atom, the 
 lower section refers to solution compared to an unmixed reference state, quoted in units of eV per carbon atom}
         \label{tab:gibbsfreeenergy}
    \end{table}
    
The solid solubility of carbon in metallic hydrogen can be calculated using the Gibbs free energies (Eqn.~\ref{eq:Gsol}). Hydrogen exhibits strong nuclear quantum effects, so our calculation includes the zero-point energy of the system, as well as the enthalpy, configurational entropy, and vibrational entropy. 
     \begin{equation}  G(P,T) = H(P,T) + U_{ZPE} + TS_{con} + TS_{vib} \label{eq:Gsol}\end{equation}     

\begin{figure}[h!]
     \centering
        \includegraphics[width=13cm]{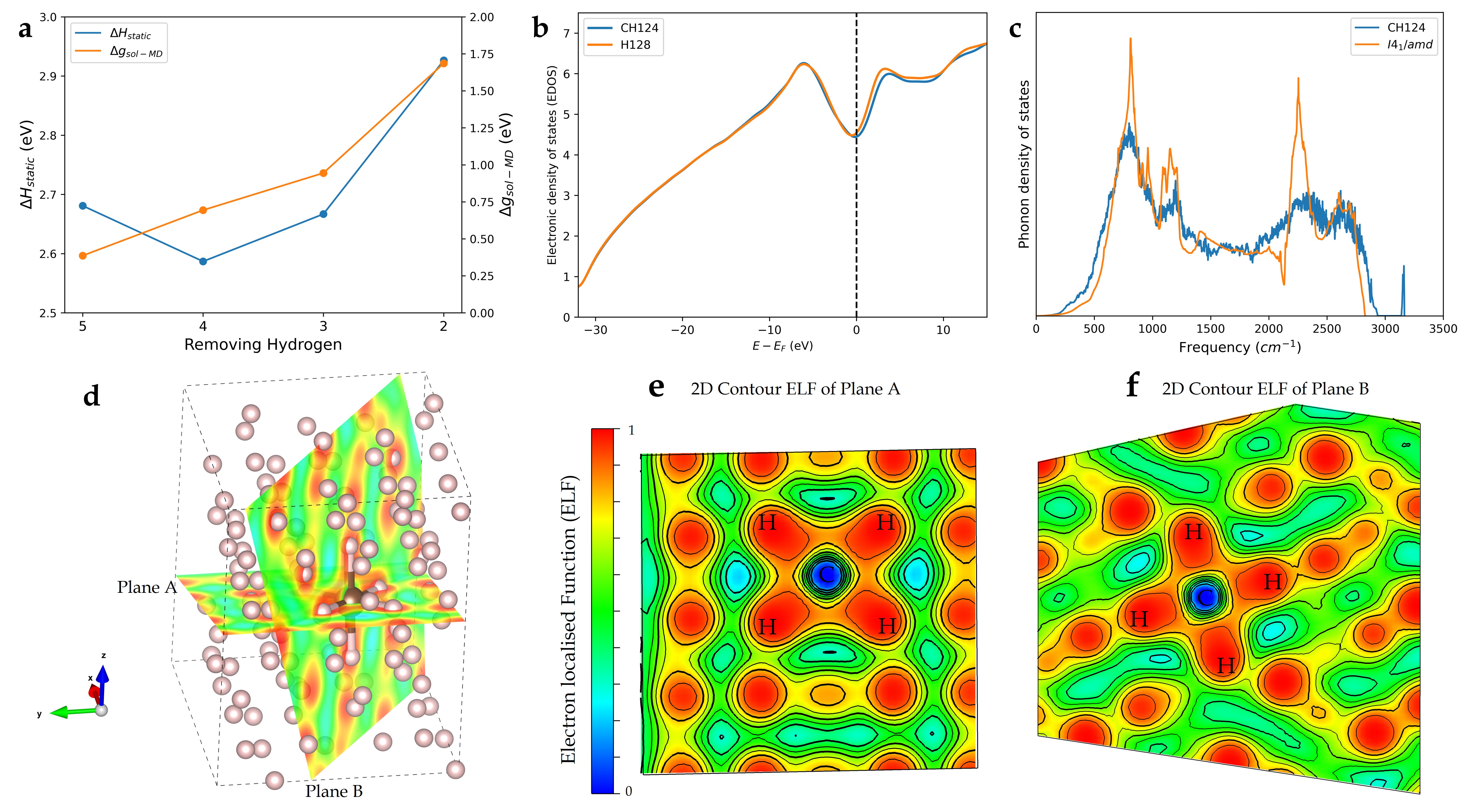} 
		\caption{\textbf{Enthalpy and Gibbs Free Energy differences, Electronic Density of States, and Electron Localised Function (ELF) of carbon in solid, metallic hydrogen.} \\
        \textbf{a)}  Enthalpy of solution (eV/atom) as a function of number of removed hydrogens from static relaxation (blue) and NPT-ensemble molecular dynamics (orange) \textbf{b)}  Electronic density of states from a typical BOMD snapshot, showing the characteristic free electron form with a pseudogap at the Fermi Energy. \textbf{c)} Phonon density of states from DFPT  of a structure from a relaxed  CH$_{124}$ MD snapshot (blue), compared with pure  $I4_1 /amd$ hydrogen (orange). \textbf{d}-\textbf{f)}  Electron Localization Function (ELF) surrounding a single carbon atom, including cross-sections of planes A and B. 
        The octahedral arrangement of the high ELF values indicates the presence of six covalent-type bonds per carbon.}
		\label{fig:free-energy-plot}
	\end{figure}

Initially we consider a single carbon solute in hydrogen.  This is more complicated than standard solid solubility calculation\cite{malerba2010ab,hepburn2013first,hepburn2015transition} because carbon is significantly larger than hydrogen, so substituting one  carbon atom for a single hydrogen is not the most stable arrangement.  We tried removing clusters of up to eight hydrogens and found that the most stable arrangement involves removing five hydrogens (Table ~\ref{tab:gibbsfreeenergy}, Fig.~\ref{fig:free-energy-plot}.)   As a check, particularly of anharmonic effects, we also calculated the free energy from the phonon density of states derived from the velocity autocorrelation function of a molecular dynamics calculation.  The results are is very good agreement for carbon, and reasonably good for hydrogen: this is expected from the different anharmonicity in the two systems.

From both MD and static enthalpy, it shows that classical enthalpy prefers pure substances to the mixture with positive values $\Delta H_{\text{MD}}$ and $\Delta H_{\text{Static}}$.  On the other hand, the entropy and zero-point energy\cite{dickey1969computer,heino2007dispersion,kong2011phonon}, favour the mixture, with negative values in both $-T\Delta S$ and $\Delta U_{ZPE}$ (see Table~\ref{tab:gibbsfreeenergy} and  Figure~\ref{fig:free-energy-plot}a).

For a single substitutional carbon, the lowest free energy  has an impurity formation energy of $\Delta g= 0.39\pm 0.17$eV, which implies a solid solubility in the parts per million range at room temperature. If more than five hydrogens are removed from the starting configuration, a vacancy defect is created which diffuses through the lattice. 
 
Strikingly, we always observe an octahedral arrangement of six neighbouring hydrogens, forming CH$_6$. The overall electronic density of states (Fig.\ref{fig:free-energy-plot}) is uninformative, being dominated by characteristic free-electron form, with the hydrogen atoms arranged in a way to produce a pseudogap at the Fermi energy, similar to isostructural Cs-IV \cite{ackland2004origin,falconi2006ab}. However, the electron localization function (ELF) around the carbon  (Figure~\ref{fig:free-energy-plot}d-f) shows that electrons are localized between the hydrogens and carbon in CH$_6$, but not between hydrogen pairs.  This suggests that carbon has reacted to form a CH$_6$ molecule with six covalent CH bonds, which we call hypermethane.

The phonon calculations show that pure hydrogen exhibits the expected two-peaked acoustic and optical branches of I4$_1$/amd.  The CH$_{124}$ supercell retains smeared-out versions of these modes, and the heavier carbon reduces the frequency of the acoustic modes in the supercell.  Strikingly, the CH$_6$ molecule has a well-defined vibrational mode at  3163 $cm^{-1}$ a higher frequency than anything in the pure metallic hydrogen (Figure~\ref{fig:free-energy-plot}c). This mode results from the in-phase vibration of the six hydrogen of CH$_6$.  Other modes involving asymmetric CH stretches are mixed with the highest frequency I4$_1$/amd modes.
    
The solid solution gives us an indication that, like silicon on earth, carbon will go from fourfold to sixfold coordination in giant planets.  However, the positive heat of solution  suggests that tempertures well above the melting point would be needed for significant solubility. We investigate this further in the next section.

\subsection{Organic compounds in liquid metallic hydrogen}
    \begin{figure}[h!]
     \centering
		\includegraphics[width=10cm]{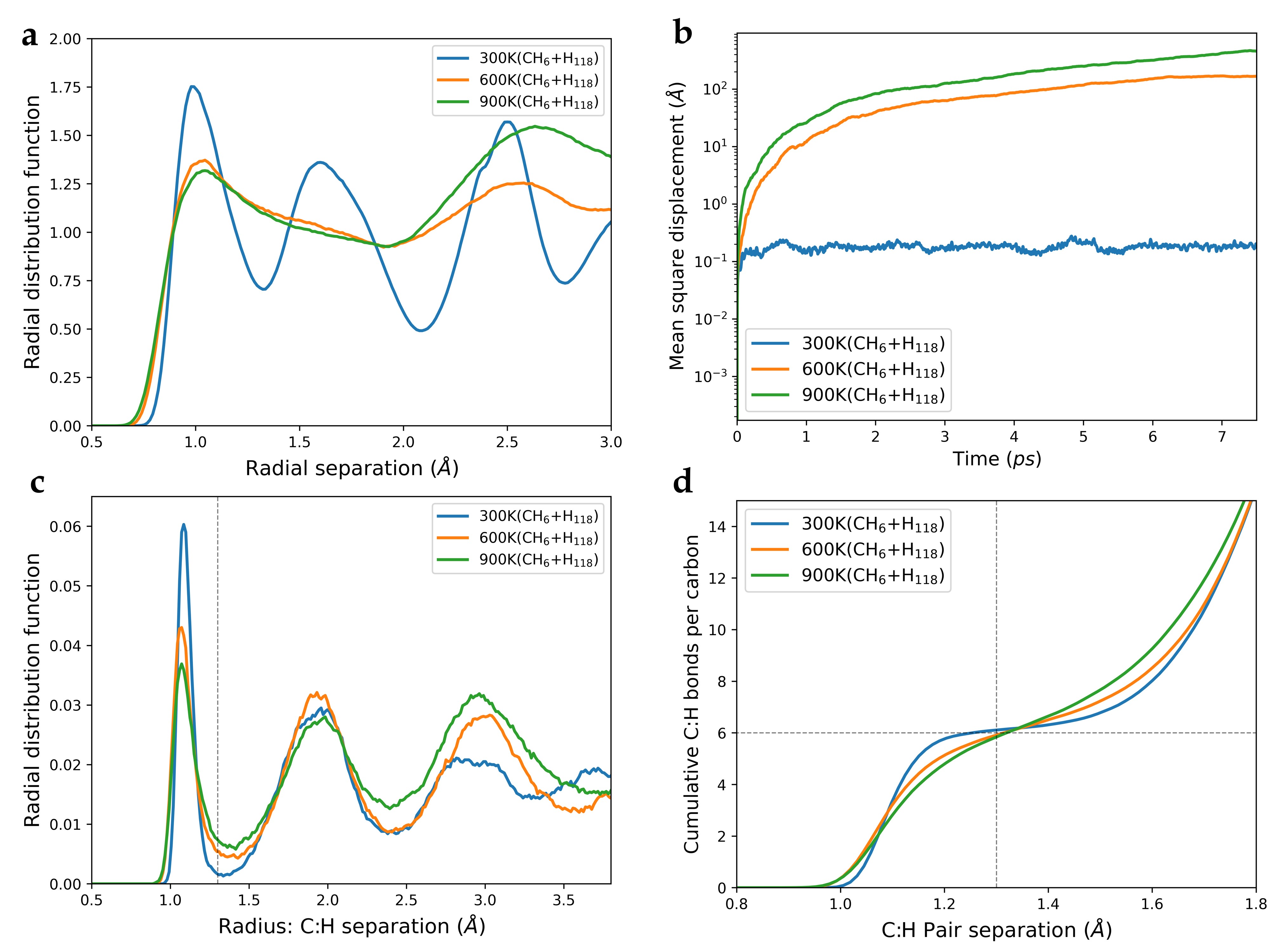}  
		\caption{\textbf{Radial Function Distribution, Mean Square Displacement, and Cumulative Number of Bonds per Carbon from BOMD in NPT ensemble.} \\
        \textbf{a)} The radial distribution function of the CH$_{6}$+H$_{118}$  BOMD simulations under the NPT ensemble at varying temperatures, and  500GPa. The blue solid line denotes the presence of structural peaks in CH$_{6}$+H$_{118}$, while the orange and green solid lines signify the disappearance of these peaks due to system melting. \textbf{b)} Mean square displacement (MSD) of CH$_{6}$+H$_{118}$ is examined, with the MSD depicting the crystal structure of CH$_{6}$+H$_{118}$ with a finite MSD at 300K, represented by the blue solid line. Conversely, cases of melting at 600K and 900K exhibit increased MSD. \textbf{c)} The radial distribution function illustrates the distribution of carbon-hydrogen pairs. The first peak of the CH-pair radial distribution function experiences smearing from 1.0 \AA~to 1.3 \AA. \textbf{d)} Cumulative number of bonds (CH bonds per carbon) is evaluated, indicating the count of hydrogens surrounding each carbon atom.}
		\label{fig:cdf-rdfmd-CH124-NPT}
	\end{figure} 

To further investigate this hypermethane, we simulated a CH$_{6}$+H$_{118}$  supercell in the NPT ensemble at 500 GPa. The radial distribution function (RDF), as shown  Figure~\ref{fig:cdf-rdfmd-CH124-NPT}(a), indicates that at 300 K, we have well-defined H-H peaks representing the I4$_1$/amd crystal. In the cases of 600 K and 900 K, the structure melted as expected\cite{geng2016predicted}, indicated by the smoothness of the RDFs in  Figure~\ref{fig:cdf-rdfmd-CH124-NPT}a). The mean square displacement (MSD)  confirms melting, with a stable MSD at 300 K and a linear increase for 600 K and 900 K, as shown in  Figure~\ref{fig:cdf-rdfmd-CH124-NPT}(b). 

In all cases, the carbon-hydrogen  RDF (Figure~\ref{fig:cdf-rdfmd-CH124-NPT}c), has a strong peak between 1.0\AA\, and 1.3\AA. Even in melted conditions, we observe hydrocarbon CH$_{6}$  molecules, as shown in  Figure~\ref{fig:cdf-rdfmd-CH124-NPT}(d), where the cumulative number of bonds in the first peak of the RDF being six. These results suggest the existence of the hypermethane CH$_{6}$  molecule above the melting line of metallic hydrogen at 500 GPa.

    \subsection{Other hypermolecules: CH$_{6}$, C$_{2}$H$_{8}$ C$_{3}$H$_{10}$, H$_{3}$O, NH$_4$ and  CH$_{4}$OH}
    
We now investigate whether more complex organic molecules can form in liquid metallic hydrogen, using NVT molecular dynamics at around 500GPa and 600K.   
We investigate five cases, adding a single carbon or oxygen atom, C$_2$ dimer, C$_3$  trimer and CO molecule to liquid metallic hydrogen. The ensemble-averaged enthalpic part of the solubility $\Delta H$ in the liquid, as shown in the table below,  suggest that in terms of enthalpy, the mixing is favoured, and more strongly so than in the solid. Unfortunately,  the zero point energy contribution was calculated this in the quasiharmonic approximation for solids, but this method cannot seriously be applied to liquids. Nevertheless, our calculation implies a high solubility of carbon in solid hydrogen, and we are not aware of any metallic system where the liquid solubility is lower than the solid.  Thus we are confident in our assertion that carbon will be sufficiently soluble in metallic hydrogen and that, for planetary compositions, it will all dissolve.

\begin{table}[]
    \centering
    \begin{tabular}{ccccc}
    \hline
    Reference states & Free energy & PV & H &  \\
    \hline
C$_{54}$ & -155.01 & 10.19 & -144.82 \\
H$_{128}$ & -13.27 & 3.53 & -9.73 \\
O$_{48}$ & -432.37 & 11.70 & -420.67 \\
N$_{64}$ & -271.35 & 10.89 & -260.46 \\
    \hline
        Compounds & Free energy & PV & H & $\Delta$ H \\
    \hline
CH$_6$ in CH$_{124}$ & -14.42 & 3.57 & -10.85 & -4.07 \\
C$_2$H$_8$ in C$_2$H$_{120}$ & -15.63 & 3.59 & -12.04 & -11.66 \\
C$_3$H$_{10}$ in C$_3$H$_{276}$ & -14.81 & 3.60 & -11.21 &  -7.30   \\
CH$_{4}$OH in COH$_{124}$ & -17.69 & 3.84 & -13.85 & 27.24 \\
OH$_3$ in OH$_{124}$& -16.70 & 3.64 & -13.06 & -4.50 \\
NH$_4$ in NH$_{124}$& -15.39 & 3.61 & -11.77 & -4.21 \\
 \hline
    \end{tabular}
    \caption{Table shows the change of enthalpy ($\Delta$H) (eV/atom) of organic compounds in liquid metallic hydrogen with reference states of carbon, oxygen, and nitrogen are diamond, monoclinic structure of C2/m space group \cite{ma2007structure}, and the helical tunnel stucture of $P2_1 2_1 2_1$ space group \cite{ma2009novel}, respectively. our calculation implies a high solubility of carbon in solid hydrogen, and we are not aware of any metallic system where the liquid solubility is lower than the solid.}
    \label{tab:my_label}
\end{table}

\begin{figure}[h]
        \centering
		\includegraphics[width=13cm]{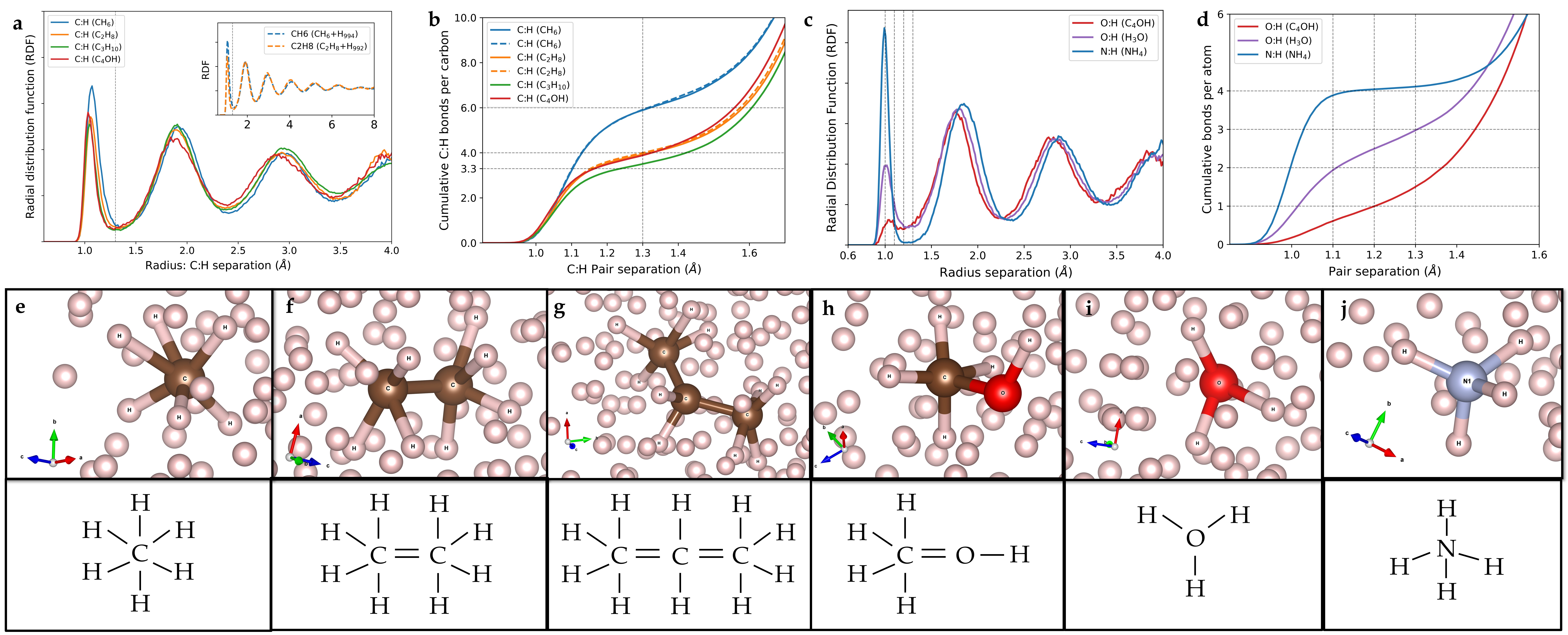}  
		\caption{\textbf{Radial Function Distribution, Cumulative Number of Bonds per Carbon, and Novel Hydrocarbons and representative snapshots from liquid simulations.} \\
        \textbf{a)} Displays the RDF of the carbon-hydrogen pairs in metallic hydrogen with one carbon (blue line), two carbons (orange line), three carbons (green line), one oxygen and a CO pair (red line), indicating a smearing of the first peak from 1.0 \AA~to 1.3 \AA. In the $\sim 10^2$ atom simulations (main figure), structure extends throughout the supercell, however the $\sim 10^3$ atom simulations (inset) indicates a suppression of these structural peaks in CH separation at greater distances. \textbf{b)} Shows the CDF (number of CH neighbours per carbon) at 1.3 \AA, revealing values of approximately 6, 4, 3.3, and 4 respectively. The short-range order is nearly identical in both system sizes (solid lines: $\sim 10^2$ atoms, dashed lines $\sim 10^3$ atoms)
        \textbf{c-d)} equivalent RDF and CDF for OH and NH pairs, showing well-defined and long-lived bonding. These values suggest the formation of new organic compounds, as depicted in \textbf{e-i)}: Snapshots from molecular dynamics showcasing novel chemical species in liquid metallic hydrogen, including \textbf{e)} CH$_6$, \textbf{f)} C$_2$H$_{8}$, \textbf{g)} C$_3$H$_{10}$, \textbf{h)} CH$_4$OH and \textbf{i)} OH$_3$ and NH$_4$, with schematic molecular bonding shown below. 
        }
		\label{fig:hydrocarbon-molecules}
\end{figure}

In each case, an exothermic reaction took place with the metallic hydrogen to produce a well-defined, stable hypermolecule with OH and CH bondlengths oscillating in the range 1.0 \AA~to 1.3 \AA. We identify these molecules as being hypervalent and hydrogen-rich  CH$_{6}$, C$_{2}$H$_{8}$ C$_{3}$H$_{10}$, NH$_4$, H$_{3}$O and  CH$_{4}$OH.  These correspond to sixfold valence for carbon, fourfold for nitrogen and trivalent oxygen, with CC and CO double bonds (Figure~\ref{fig:hydrocarbon-molecules}).  We note that this mimics the high pressure behaviour of the equivalent second row elements: hexavalent Silicon\cite{wentzcovitch1995ab} tetravalent phosphorus and trivalent Sulphur\cite{drozdov2015conventional}.

Figure~\ref{fig:hydrocarbon-molecules} shows the CH, NH and OH radial and cumulative distribution functions for the  hypermolecules we investigated, with the sharp first peak in RDF defining the covalent bond, and the plateau in CDF showing  the coordination. All hypermolecules remain stable throughout the 10ps simulation.  The covalent bonding is further evidenced by ELF analysis.  The negative charge on these hypermolecules is evidenced from molecular orbital considerations and the observation of screening with Friedel oscillations.

We can understand the  chemistry of these hypermolecules by considering that, in a metallic environment, extra electrons are readily available to stabilise charged molecules. The local electronic structure of the molecule shows a similar situation to the hypermethane solid solubility, with electrons in covalent type bonds. 
The hypermolecules carry a formal negative charge from the excess of electrons, e.g. CH$_6$ has 14 electrons and a total nuclear charge of only 12.  This charge will be screened by the surrounding liquid.
The simple case of a charge in a metal  gives slowly decaying Friedel oscillations\cite{harrison1979solid,ashcroft1976solid} in the electrostatic potential.  Fluid metallic hydrogen is a more complicated case because both protons and electrons play a role in the screening.  Our large simulations with around 1000 atoms for CH$_6$ shows that there are Friedel oscillations in the proton density extending to seven distinctive peaks.   

\section{Methods}\label{sec11}

   Calculations were performed using density functional theory (DFT) \cite{DFT-HK,DFT-KS} implemented in both Quantum Espresso (QE) \cite{giannozzi2009quantum,giannozzi2017advanced} and Born-Oppenheimer molecular dynamics (BOMD) implemented in the CAmbridge Serial Total Energy Package (CASTEP) \cite{CASTEP}.

    Using the Broyden–Fletcher–Goldfarb–Shanno algorithm (BFGS) method \cite{BFGS,liu1989limited,RMP-Payne}, the four-atom conventional $I4_{1}/amd$ structure of atomic metallic hydrogen and the diamond (for carbon) were fully optimized at 500GPa, using a force convergence criterion of $10^{-5}$ eV/\AA\ and a  very dense Monkhorst-Pack grid k mesh. We used the exchange-correlation functional of Perdew–Burke–Ernzerhof (GGA-PBE) \cite{perdew1996generalized}. The Born-Oppenheimer molecular dynamics (BOMD), time step was 0.5 fs with  velocity-Verlet integration\cite{VV}. The isothermal–isobaric ensemble (NPT) \cite{NHC} was implemented, employing the Parrinello-Rahman barostat\cite{ParRah}. Additionally, a thermostat was set at 300 K, and  Berendsen thermostat \cite{Berendsen}. Solubility calculations, both static relaxation ($a = 4.84$ \AA and $c = 6.23$ \AA) and NPT molecular dynamics, were based on a $4\times4\times2$ supercell (128 hydrogens) of the four-atom conventional $I4_{1}/amd$ structure, with carbon substituted for some hydrogens. 
    
    For the BOMD calculations in the melt, our long simulations were run with boxes containing CH$_{124}$, C$_{2}$H$_{120}$, and C$_{3}$H$_{276}$, COH$_{124}$, OH$_{124}$, and NH$_{124}$ respectively. The CH$_{6}$, C$_2$H$_{10}$,  CH$_4$OH,  OH$_3$, and NH$_{4}$ molecule simulations were initiated using a $4\times4\times2$ supercell with lattice constants $a = 4.84$~\AA~and $c = 6.23$~\AA~where four hydrogens are removed for each inserted oxygen, carbon, or nitrogen .  For C$_3$H$_8$, a $6\times6\times2$ supercell (288 hydrogens) with lattice constants $a = 7.25$~\AA~ and $c = 6.23$~\AA~ was used. The simulations of melted samples used the NVT ensemble.  In the carbon BOMD-NPT simulation, we used a $3\times3\times3$ supercell of the two-atom primitive cell of diamond with the same parameters as the other cases. All of our BOMD simulations were performed up to 20,000 steps and checked for convergence of potential energy.  In no case did the hypermolecules dissociate.

    For lattice dynamics, we analyzed the phonon spectrum of the diamond structure and the four-atom conventional $I4_{1}/amd$ structure of atomic metallic hydrogen under a pressure of 500 GPa using Quantum ESPRESSO (QE). Structural optimization was conducted using the BFGS method \cite{BFGS,liu1989limited}, fully relaxing crystal structures with a force convergence criterion of $1.0^{-5}$ eV/\AA. The Monkhorst-Pack grid k mesh \cite{monkhorst1976special} employed a dense grid, with Marzari-Vanderbilt-DeVita-Payne cold smearing of 0.02 Ry applied to the Fermi surface \cite{marzari1999thermal}. As with CASTEP, the PBE\cite{perdew1996generalized}, exchange correlation energy functional was implemented, and we used optimized norm-conserving Vanderbilt pseudopotentials \cite{hamann2013optimized,schlipf2015optimization}.  The lattice dynamics was performed using QE based on density functional perturbation theory (DFPT) \cite{baroni2001phonons}.  The electronic density of states of CH$_{124}$ and H$_{128}$ were computed with the optimized structures from the final step of the BOMD simulations.

We find the solubility limit from equating the Gibbs free energy in the mixture with that in the pure substances I$4_1$/amd hydrogen and diamond carbon at 500GPa\cite{clark1995theoretical}. 
    \begin{equation}  G_{xy}(P,T) + k_BT\left [c\ln c + (1-c)\ln (1-c) \right ] = xG_H(P,T)+ yG_C(P,T)  \label{eq:sol}\end{equation}
    Where $c=y/(x+y)$ is the carbon concentration and $y=1-x$ is the atomic fraction of C.

For the liquid simulations we use the NVT ensemble at density and system-sizes equivalent to the I4$_1$/amd.  We  compare these with larger simulations of around 1,000 atoms which previous work \cite{geng2019thermodynamic} has shown sufficient to converge the RDF of metallic liquid hydrogen.

Our initial analysis is based on partial  RDFs of  CH and OH separations, as shown in (a) of Figure~\ref{fig:hydrocarbon-molecules}.  This shows that  simulations of around one hundred or one thousand atoms gives the same hypermolecule formation, both number of bonds and bondlength Figure~\ref{fig:hydrocarbon-molecules}(b).  We observe liquid structure peaks which extend beyond the small unit cell size,  however, within a thousand-atom simulation,  this oscillating structure has decayed away exponentially  (Figure~\ref{fig:hydrocarbon-molecules}a).  Therefore, we are confident that the results from our hundred-atom simulations give a good description of the hypermolecules.

\section{Discussion and Conclusions}\label{sec13}

In this study, we used density functional theory calculation to predict the existence of hyperorganic molecules such as CH$_6$, C$_2$H$_8$, C$_3$H$_{10}$, OH$_3$,  NH$_4$ and CH$_4$OH in metallic hydrogen.   OH$_3^+$ and NH$_4^+$ are well-known cations: hexavalent carbon  CH$_6^{2+}$ is a novelty  

The chemical bonding can be understood in terms of standard covalent chemistry, with hypervalent carbon/oxygen forming single bonds to H, and double bonds between heavier elements. This means that these hyperorganic molecules are negatively charged: the charges are screened by the surrounding metallic hydrogen and we show that this causes an oscillating charge density wave around the molecules.

Regarding consequences for astronomy, exact details will vary from planet to planet, depending on its history.  We note that the temperatures considered here are considerably lower than in the cores of Jupiter and Saturn, but well about the equivalent blackbody temperature of these planets \cite{wallace2006atmospheric}.

The positive heat of solution for carbon suggests that  carbon will condense and fall as diamond rain, again depending on the gravitational field as well as the chemistry. However, under the conditions expected in gas giants with metallic hydrogen cores this solubility limit is more than parts per thousand, close to or above the expected primordial carbon-hydrogen ratio.   Therefore, we anticipate that in many gas giant planets a significant proportion of the carbon will remain in solution in metallic hydrogen.   

Terrestrial synthesis of these molecules is challenging but within the reach of current methodology.  Liquid metallic hydrogen forms at lower pressures than its solid counterpart, in both static and dynamic compression \cite{zaghoo2017conductivity,zaghoo2018striking,knudson2015direct}. So creation of hypermolecules is plausible, but detection is more difficult.  Our CH$_6$ calculation suggests that the molecules will have distinctive vibrational modes beyond the atomic hydrogen frequencies, but measuring these in such an extreme, metallic environment will be challenging
 
Since all our simulations produces long-lived hypermolecules, it seems certain that more complex molecules will also be stable.   Thus it appears that the metallic hydrogen environment, the most common state of condensed matter in the universe,  is capable of supporting its own rich organic chemistry.

\backmatter

%\bmhead{Supplementary information}

%If your article has accompanying supplementary file/s please state so here. 

%Authors reporting data from electrophoretic gels and blots should supply the full unprocessed scans for key as part of their Supplementary information. This may be requested by the editorial team/s if it is missing.

%Please refer to Journal-level guidance for any specific requirements.

\bmhead{Acknowledgements}
This research project is supported by the Second Century Fund (C2F), Chulalongkorn University. GJA acknowledges funding from the ERC project Hecate. This work used the Cirrus UK National Tier-2 HPC Service at EPCC (http://www.cirrus.ac.uk) funded by the University of Edinburgh and EPSRC (EP/P020267/1). This also work used the ARCHER2 UK National Supercomputing Service (https://www.archer2.ac.uk) as part of the UKCP collaboration. We acknowledge the supporting computing infrastructure provided by NSTDA, CU, CUAASC, NSRF via PMUB [B05F650021, B37G660013] (Thailand). URL:www.e-science.in.th. We thank Pattanasak Teeratchanan, David Ceperley and Jeffrey M. McMahon for their valuable suggestions on DFT(QE)-related issues for studying metallic hydrogen. We thank Miriam Pena-Alvarez and Stewart McWilliams for comment and proofreading.

    \section*{Data availability statement}
	The data cannot be made publicly available upon publication because they are not available in a format that is sufficiently accessible or reusable by other researchers. The data that support the findings of this study are available upon reasonable request from the authors.

    \section*{Conflict of interest}
	The authors have no conflicts of interest to declare. All co-authors have seen and agree with the contents of the manuscript and there is no financial interest to report. We certify that the submission is original work and is not under review at any other publication.

\bibliographystyle{}
\bibliography{sn-bibliography}
\end{document}